\documentclass[preprint]{aastex}

\usepackage{natbib}
\newcommand{\tten}{\times 10}
\newcommand{\myemail}{ysy@umd.edu}

\newcommand{\nimpra}{Nucl. Instrum. Meth. A}

\newcommand{\icrc}{Int. Cosmic-Ray Conf.~}
\newcommand{\asr}{Adv. Spa. Res.}
\newcommand{\astp}{Astropart. Phys.}
\newcommand{\plb}{Phys. Lett. B}
\newcommand{\npbp}{Nucl. Phys. B-Proc. Sup.}
\newcommand{\zphyc}{Z. Phys. C Part. Fields}

\newcommand{\brasp}{B. Russ. Acad. Sci. Phys.}
\newcommand{\ncc}{Nuovo Cimento C}

\begin{document}

\title{Cosmic-Ray Proton and Helium Spectra\\from the First CREAM Flight}

\author{
Y.~S. Yoon\altaffilmark{1,2*},
H.S. Ahn\altaffilmark{1},
P.~S. Allison\altaffilmark{3}, 
M.~G. Bagliesi\altaffilmark{4},
J.~J. Beatty\altaffilmark{3},
G. Bigongiari\altaffilmark{4}, 
P.~J. Boyle\altaffilmark{5\dag}, 
J.~T. Childers\altaffilmark{6\ddag},
N.~B. Conklin\altaffilmark{7\ddag\dag},
S. Coutu\altaffilmark{7}, 
M.~A. DuVernois\altaffilmark{6},
O. Ganel\altaffilmark{1}, 
J.H. Han\altaffilmark{1},
J.A. Jeon\altaffilmark{8},
K.C. Kim\altaffilmark{1},
M.H. Lee\altaffilmark{1},
L. Lutz\altaffilmark{1},
P. Maestro\altaffilmark{4},
A. Malinine\altaffilmark{1},
P.S. Marrocchesi\altaffilmark{4},
S.~A. Minnick\altaffilmark{9},
S.~I. Mognet\altaffilmark{7\ddag\ddag},
S. Nam\altaffilmark{8},
S. Nutter\altaffilmark{10},
I.~H. Park\altaffilmark{8},
N.H. Park\altaffilmark{8,5},
E.S. Seo \altaffilmark{1,2},
R. Sina\altaffilmark{1},
S. Swordy\altaffilmark{5},
S.~P. Wakely\altaffilmark{5}, 
J. Wu\altaffilmark{1},
J. Yang\altaffilmark{8},
R. Zei\altaffilmark{4},
S.Y. Zinn\altaffilmark{1\S} }

\altaffiltext{1}{Institute of Physical Science and Technology, University of Maryland, College Park, MD 20742, USA}
\altaffiltext{2}{Department of Physics, University of Maryland, College Park, MD 20742, USA}
\altaffiltext{3}{Department of Physics, Ohio State University, Columbus, OH 43210, USA}
\altaffiltext{4}{Department of Physics, University of Siena and INFN, Via Roma 56, 53100 Siena, Italy}
\altaffiltext{5}{Enrico Fermi Institute and Department of Physics, University of Chicago, Chicago, IL 60637, USA}
\altaffiltext{6}{School of Physics and Astronomy, University of Minnesota, Minneapolis, MN 55455, USA}
\altaffiltext{7}{Department of Physics, Penn State University, University Park, PA 16802, USA}
\altaffiltext{8}{Department of Physics, Ewha Womans University, Seoul 120\textendash750, Republic of Korea}
\altaffiltext{9}{Department of Physics, Kent State University, Tuscarawas, New Philadelphia, OH 44663, USA}
\altaffiltext{10}{Department of Physics and Geology, Northern Kentucky University, Highland Heights, KY 41099, USA}
\altaffiltext{*}{Email: \myemail}
\altaffiltext{\dag}{Now at McGill University, Montr\'{e}al, Qu\'{e}bec H3A 2T8, Canada}
\altaffiltext{\ddag}{Now at Universit\"{a}t Heidelberg, Kirchhoff-Institut f\"{u}r Physik, 69120 Heidelberg, Deutschland}
\altaffiltext{\ddag\dag}{Now at Gannon University, Erie, PA 16541, USA}
\altaffiltext{\ddag\ddag}{Now at Department of Physics and Astronomy, University of California, Los Angeles, CA 90095, USA}
\altaffiltext{\S}{Now at Promogen, Inc., Rockville, MD 20850, USA}

\begin{abstract}
Cosmic-ray proton and helium spectra have been measured with the balloon-borne
Cosmic Ray Energetics And Mass experiment flown for 42 days in Antarctica
in the 2004\textendash2005 austral summer season.
High-energy cosmic-ray data were collected at an average altitude of $\sim$38.5 km
with an average atmospheric overburden of $\sim$3.9 g cm$^{-2}$. Individual elements
are clearly separated with a charge resolution of $\sim$0.15 $e$ (in charge units) 
and $\sim$0.2 $e$ for protons and helium nuclei, respectively.
The measured spectra at the top of the atmosphere are represented by power laws
with a spectral index of $-$2.66 $\pm$ 0.02 for protons from 2.5 TeV to 250 TeV and
\textendash2.58 $\pm$ 0.02 for helium nuclei from 630 GeV nucleon$^{-1}$ to 63 TeV nucleon$^{-1}$.
They are harder than previous measurements at a few tens of GeV nucleon$^{-1}$.
The helium flux is higher than that expected from the extrapolation of the power law 
fitted to the lower-energy data.
The relative abundance of protons to helium nuclei is 9.1 $\pm$ 0.5
for the range from 2.5 TeV nucleon$^{-1}$ to 63 TeV nucleon$^{-1}$.
This ratio is considerably smaller than the previous measurements
at a few tens of GeV nucleon$^{-1}$.

\end{abstract}


\section{Introduction}

Cosmic rays are the product of energetic processes in the universe, and their interactions 
with matter and fields are the source of much of the diffuse gamma-ray, X-ray, 
and radio emissions that are observed.
Therefore, the origin of cosmic rays and how they propagate have
a major impact on our understanding of the universe.
Supernova shock waves could provide the 
power required to sustain the galactic cosmic-ray intensity,
but details of the acceleration mechanism are not completely understood.
The shock acceleration mechanism is believed to be a prevalent process 
in astrophysical plasmas on all scales throughout the universe. 
It has been shown to work in the heliosphere, e.g., at planetary bow shocks, 
at interplanetary shocks in the solar wind, and at the solar wind termination shock. 

It is a characteristic of diffusive shock acceleration that the resulting particle energy spectrum 
is much the same for a wide range of shock properties. 
This energy spectrum, when corrected for leakage from the Galaxy, is consistent with 
the observed spectrum of Galactic cosmic rays. In the most commonly used form of the theory, 
the characteristic limiting energy is about $Z$ $\times$ $10^{14}$ eV, 
where $Z$ is the particle charge \citep{Lagage1983}.
The observed composition should begin to change beyond about $10^{14}$ eV, 
the limiting energy for protons, and the Fe spectrum would start to steepen 
at an energy 26 times higher. In this scenario, protons would be the most dominant element 
at low energies, but heavier elements would become relatively more abundant at higher energies, 
at least up to the acceleration limit for iron. 

Compelling evidence that supernova remnants (SNRs) are common sites for shock acceleration of electrons 
comes from observations of non-thermal synchrotron radiation from several shell-type remnants 
\citep{Koyama1995, Allen1997, LeBohec2000}. 
Non-thermal X-ray spectra indicate the presence of very high energy electrons which, 
at least in the case of SN 1006, have energies $>$2 $\times 10^{14}$ eV \citep{Koyama1995}.
These electrons were likely accelerated at the remnant because at this energy electrons 
cannot travel far from their origin before they are attenuated by synchrotron losses. 
There are other sources of particle acceleration that may also contribute to 
the cosmic-ray beam \citep{Dermer2001}.
Recent \textit{Chandra} X-ray observations of \textit{Tycho's} SNR have shown 
hot stellar debris keeping pace with an outward-moving shock wave indicated 
by high-energy electrons. 

Semi-direct evidence for the acceleration of cosmic-ray protons could come 
in the form of gamma rays from pion decay \citep{Ellison2005}.
Indeed, the observation of TeV gamma rays, possibly of $\pi^{0}$-origin,
from the SNR RX J1713.7$-$3946 \citep{Enomoto2002,Aharonian2007} 
may have revealed the first specific site where protons are accelerated 
to energies typical of the main cosmic-ray component. 
Their hadronic origin is yet to be confirmed, 
but the CANGAROO collaboration has shown that the energy spectrum of gamma-ray emission 
from SNR RX J1713.7$-$3946 matches that expected if the gamma rays 
are the decay products of neutral pions generated in \textit{p-p} collisions.
 Although the proton scenario is favored because of the spectral shape, 
gamma rays may originate from either electrons or protons. 
A complete understanding of gamma-ray emission processes may need a broadband approach 
\citep{Aharonian2006}, 
using all the available measurements in different wavelength regions. 
Direct measurements of nuclear particle composition changes would provide 
strong corroborating evidence that shocks associated with shell-type SNRs 
provide the acceleration sites for cosmic rays. 
 
Shock acceleration is the generally accepted explanation for the characteristic power-law feature 
of cosmic-ray energy spectra, although ground-based measurements have shown that 
the all-particle spectrum extends far beyond the highest energy thought possible 
for supernova shock acceleration. 
These measurements have also shown that the energy spectrum above $10^{16}$ eV is 
somewhat steeper than the spectrum below $10^{14}$ eV, which lends credence 
to the possibility of a different source. 
Of course, the ``knee'' structure might be related to energy-dependent leakage effects 
during the propagation process \citep{Ptuskin1993,Swordy1995} or to other effects, 
such as reacceleration in the galactic wind \citep{Voelk2003} and acceleration 
in pulsars \citep{Bednarek2002}. 
Whether and how the spectral ``knee'' is related to the mechanisms of 
acceleration, propagation, and confinement are among the major current questions 
in particle astrophysics. 

\section{CREAM Experiment}

The Cosmic Ray Energetics And Mass (CREAM) experiment \citep{Seo2008} was designed and constructed
to extend balloon and space-based direct measurements of cosmic-ray elemental spectra
to the highest energy possible in a series of balloon flights.
The detailed energy dependence of elemental spectra at very high energies,
where the rigidity-dependent supernova acceleration limit could be reflected in composition change,
provides a key to understanding the acceleration and propagation of cosmic rays. 
We report in this paper the proton and helium spectra as well as their ratios observed
from the maiden flight of the CREAM payload in Antarctica.
Results from the CREAM experiment such as B/C ratio and heavy elemental spectra
are discussed in elsewhere
\citep{Ahn2008TRD,Ahn2009CREAM2,Ahn2010CREAM12}.

\subsection{CREAM Flight 2004\textendash2005}

The first Long Duration Balloon (LDB) flight of the CREAM payload was launched from McMurdo Station,
Antarctica on December 16, 2004.
It subsequently circumnavigated the South Pole three times
for a record-breaking duration of 42 days; the flight was terminated on January 27, 2005.
The instrument float altitude remained between 37 and 40 km through most of the flight.
The corresponding atmospheric overburden was 3.9 $\pm$ 0.4 g cm$^{-2}$.
The diurnal altitude variation due to the Sun angle change was very small, $<$1 km, near the pole,
i.e., at high latitude, which increased as the balloon spiraled out to lower latitudes \citep{Seo2008}.
The temperature of the various instrument boxes stayed within the required
operational range with daily variation of a few $^{\circ}$C, consistent with the Sun angle. 
A total of 60 GB of data including $\sim$4 $\times$ 10$^{7}$ science events were collected. 

The science instrument was supported by the command and data module developed
by the NASA Wallops Flight Facility (WFF) \citep{Thompson2008}.
This is in contrast to typical LDB payloads which utilize the support instrumentation package 
provided by the Columbia Scientific Balloon Facility.
CREAM was the first LDB mission to transmit all the prime science and housekeeping data (up to 85 kbps)
in near real-time through the Tracking and Data Relay Satellite System (TDRSS) via a high-gain antenna,
in addition to having an onboard data archive.
To fit the data into this bandwidth, science event records excluded information
from channels that had levels consistent with their pedestal value.
This ``data sparsification'' reduced the average high-energy shower event record size by nearly 95$\%$. 
The science instrument was controlled from a science operation center at the University of Maryland
throughout the flight after line-of-sight operations ended at the launch site. 
Primary command uplink was via TDRSS, with Iridium serving as backup whenever the primary link
was unavailable due to schedule or traversing zones of exclusion.
The nearly continuous availability of command uplink and data downlink allowed rapid response
to changing conditions on the payload (e.g., altitude-dependent effects) throughout the flight. 
More details about flight operations and the data acquisition system are discussed elsewhere
\citep{Yoon2005Op,Zinn2005}.

\subsection{CREAM Instrument}

\begin{figure} [t]
\epsscale{.60}
\plotone{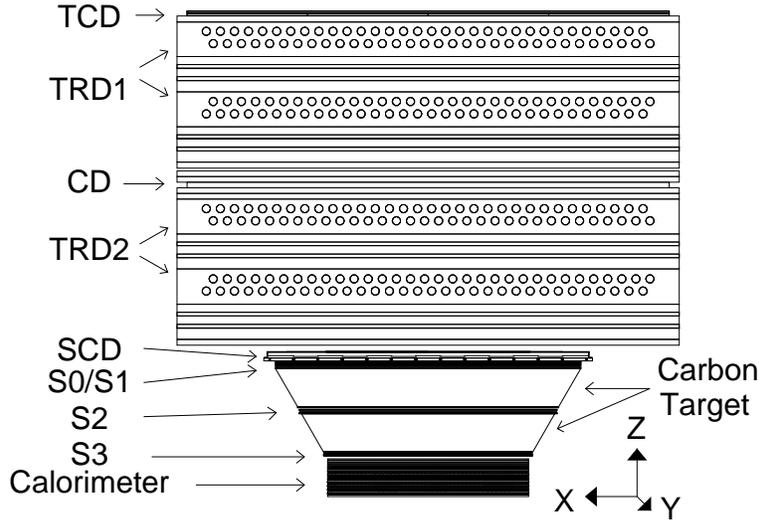}
\caption{CREAM detector configuration. 
From the top are shown the TCD, upper TRD1, CD, lower TRD2, SCD, Hodoscope S0/S1, carbon targets, 
S2, S3 and the calorimeter.\label{fig:inst}}
\end{figure}

The instrument was designed to meet the challenging and conflicting requirements
to have a large enough geometry factor to collect adequate statistics for the low flux
of high-energy particles, and yet stay within the weight limit for near-space flights \citep{Ahn2007CREAMInst}. 
It was comprised of a suite of particle detectors to determine the charge
and energy of the very high energy particles. 
As shown schematically in Figure \ref{fig:inst}, the detector configuration
included a timing charge detector (TCD), a transition radiation detector (TRD)
with a Cerenkov detector (CD), a silicon charge detector (SCD), hodoscopes (HDS),
and a tungsten/scintillating fiber calorimeter.
Starting from the top, the TCD consists of two crossed layers of four 5 mm thick
and 1.2 m long plastic scintillators \citep{Ahn2009TCD}.
It defines the 2.2 m$^{2}$ sr trigger geometry and
determines charge based on the fact that the incident particle enters the TCD
before developing a shower in the calorimeter, and the backscattered albedo particles
arrive several nanoseconds later.
A layer of scintillating fibers, S3, located between the carbon target and
the tungsten calorimeter provides a reference time.  

The TRD determines the Lorentz factor for $Z$ $\geqslant$ 3 nuclei by measuring transition X-rays
using thin-wall gas tubes.
Transition radiation is produced when a relativistic particle traverses an inhomogeneous medium,
in particular the boundary between materials of different dielectric properties.
The TRD consists of a foam radiator and 16 layers of proportional tubes filled
with a mixture of xenon (95$\%$) and methane (5$\%$) gas \citep{Ahn2008TRD}.
The CD between the two TRD sections provides low-energy particle rejection
at the flight site, Antarctica, where the geomagnetic cutoff is low. 
It also provides additional charge identification.

The SCD is comprised of 380 $\mu$m thick Si sensors \citep{Park2007}.
It is segmented into pixels, each about 2.12 cm$^{2}$ in area
to minimize multiple hits in a segment due to backscattered particles.
The targets are comprised of blocks of densified graphite cemented in carbon/epoxy composite cradles.
The vertical thickness of the carbon targets is about 0.5 interaction lengths.
They force hadronic interactions in the calorimeter, which measures the shower energy
and provides tracking information to determine which segment(s) of the charge detectors
to use for the charge measurement \citep{Seo1996}.   

The calorimeter consists of 20 tungsten layers interleaved with scintillating fiber ribbon layers
which are alternately oriented in the $x$- and $y$-directions. 
Each tungsten layer is 1 radiation length thick to sample the shower every radiation length.
Each layer consists of fifty 1 cm wide and 0.5 mm thick fiber ribbons
to measure the longitudinal and lateral distributions of the shower.
The light signal from each ribbon is collected by means of an acrylic light-mixer coupled
to a bundle of clear fibers.
This is split into three sub-bundles, each feeding a pixel of a hybrid photo diode (HPD).
In this way the wide dynamic range of the calorimeter is divided into three sub-ranges
(low, mid, high) with different gains, chosen to match the dynamic range of the front-end electronics \citep{Lee2006}.
Tracking for showers is accomplished by extrapolating each shower axis back to the charge detectors.
The HDS S0/S1 and S2, comprised of 2 mm thick and 2 mm wide scintillating fibers,
provide additional tracking information above the tungsten stack \citep{Yoon2005HDS,Marrocchesi2004}. 
The tracking uncertainty is smaller than the pixel size of the SCD \citep{Ahn2001}.

Tracking for non-interacting particles is achieved in the TRD with better accuracy
(1 mm resolution with 67 cm lever arm, 0.0015 radians). 
The TRD and calorimeter have different systematic biases in determining particle energy.
The use of both instruments allows in-flight cross-calibration of the two techniques and,
consequently, provides a powerful method for measuring cosmic-ray energies \citep{Maestro2007}. 
Details of the detectors and their performance are discussed elsewhere
\citep{Lee2006,Park2004,Ahn2007CREAMInst}.  

\section{Data Analysis}

The main trigger conditions for science events were
(1) significant energy deposit in the calorimeter for high-energy particles or
(2) large pulse height, $Z$ $>$ 2, in the TCD for heavy nuclei. 
The former requires each of six consecutive layers in the calorimeter 
to have at least one ribbon recording a deposit of more than 45 MeV.
The high-energy shower events that meet this calorimeter trigger condition were used in this analysis.  

\subsection{Event Selection}
\label{sec:evtsel}
 
The ribbon with the highest energy deposit and the neighboring ribbons
on both sides were used to determine the position in each layer of maximum energy deposits.
The shower axis was reconstructed by a least-squares fit of a straight line
through a combination of these hit positions in the $XZ$ and $YZ$ planes \citep{Ahn2007CREAMHeavy}.
Hits not along the straight line were excluded from the fit.
The resulting trajectory resolution is $\sim$1 cm when projected to the SCD.
The reconstructed trajectories were required to traverse the SCD active area
and the bottom of the calorimeter active area.

At this stage non-interacting particles are removed,
but some events have their first hadronic interaction in the calorimeter layers 
instead of the carbon targets.
These late interacting events could result in an underestimation of deposited energy,
or misidentification of charge due to large uncertainties in the trajectory reconstruction.
Since their longitudinal shower profiles are different, events
with small energy deposit in the top few layers of the calorimeter
were removed to ensure that the selected events had their first interactions
either in the carbon targets or in the top of the calorimeter. 

\subsection{Charge Determination}

\begin{figure} [t]
\epsscale{.45}
\plotone{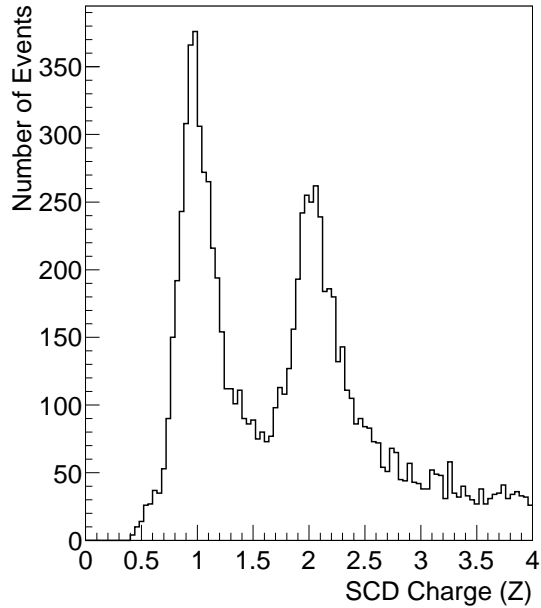}
\caption{Charge distribution of selected events determined with the SCD.
The number of protons and helium nuclei are identified from the charge distribution 
in each energy bin. \label{fig:scdz}}
\end{figure}

In order to determine the incident particle charge, the reconstructed shower axis 
from the calorimeter
was extrapolated to the SCD and a 7 $\times$ 7 pixel area, about 10 $\times$ 10 cm$^{2}$,
centered on the extrapolated position, was scanned to seek for the highest pixel signal.
The scanned area was optimized to sustain the charge identification efficiency of 99$\%$
in all energy bins,
accounting for dead and noisy SCD and calorimeter channels ($\sim$15$\%$ and 13$\%$, respectively), 
and determined to be a 7 $\times$ 7 pixel area.
That highest pixel signal was then corrected for the particle path length
(calculated from the reconstructed incidence angle) in the sensor. 
The signal reflects the ionization energy loss per unit path length ($dE/dx$) 
of an incident particle in the SCD. 
The energy loss is proportional to $Z^{2}$. 
According to Monte Carlo (MC) simulations and beam tests, the expected contamination
from secondary particles back scattering from the calorimeter is $<$3$\%$
when this tracking-based selection method is used \citep{ParkN2007}.
The resulting SCD signal distribution is shown in Figure \ref{fig:scdz}. 
Events with $Z$ $<$ 1.7 were selected as protons, while events with 1.7 $\leqslant$ $Z$ $<$ 2.7 were selected as helium nuclei.
The charge resolutions are estimated as $\sim$0.15 $e$ and $\sim$0.2 $e$ for protons and helium nuclei, respectively.
The proton and helium losses due to $dE/dx$ Landau tails were corrected by charge selection efficiencies,
which will be discussed in Section \ref{sec:absoluteflux}. 
The proton events in the helium range were removed as a background in the helium selection, 
and the helium events in the proton range were removed as a background in the proton selection,
which will be discussed in Section \ref{sec:background}. 
Unstable SCD channels identified by their large root-mean-square pedestal variations
throughout the flight were excluded from the analysis. 
Including dead or noisy channels, $\sim$15$\%$ of the total 2,912 SCD channels were masked. 

\subsection{Energy Measurement}

An ionization calorimeter is the only practical way to measure the energy of protons and helium nuclei
above $\sim$1 TeV, but calorimeters with full containment of hadronic showers are too massive
to be incorporated into space-based or balloon-borne experiments \citep{Ganel1999}. 
A thin calorimeter offers a practical approach but the calorimeter calibration requires
the use of accelerator beam particles having known energy.
The CREAM calorimeter was calibrated before the flight with electron beams
at the European Organization for Nuclear Research (CERN).
Each of the 1000 fiber ribbons was exposed to 150 GeV electrons.
The responses from the 50 ribbons in a given layer are equalized by moving the detector
in steps of 1 cm vertically or 1 cm horizontally, so the electron beam is centered
each time on the center of a different ribbon in each X or Y layer. 

The calorimeter was designed to measure the energy deposit from showers initiated
by nuclei with energies up to 10$^{15}$ eV and higher.
Its sampling fraction for isotropically incident TeV proton showers initiated in the graphite targets
is about 0.13$\%$ of the parent's energy in the active media.
With electron test beam energies of 150 GeV or less, only 8$-$10 layers
around the shower maximum register enough scintillation to allow calibration.
To address this, the calibration scan was carried out in three sets of runs
by exposing the calorimeter from the bottom with additional targets along the beam line
as described in \citet{Ahn2007CREAMInst}.

The energy deposit expected along the shower core in each layer was calculated using
MC simulations of electron showers.
Conversion factors from analog-to-digital conversion (ADC) signals to MeV
were obtained from the ratio of MC simulation of the energy deposited in each ribbon
to the measured ADC signal from the calibration beam test.
The MC simulations were based on GEANT/FLUKA 3.21 \citep{Brun1984,Fasso1993}.
The ADC signals were corrected for the HPD quantum efficiency
and gain difference from the different HPD high-voltage settings between the beam test and the flight. 

Inter-calibration between the low- and mid- energy ranges,
and between the mid- and high- energy ranges were carried out with flight data
by comparing the signals from two ranges of the same ribbon generated by the same shower.
None of the proton and helium event candidates saturated in the middle range,
so the high range optical division was not needed for this analysis.
More details about the calibration can be found in \citet{Yoon2005BT,Yoon2007Calib} and \citet{Ahn2006BT}. 

The calorimeter, HDS, and SCD were also exposed to nuclear fragments (A/Z = 2)
of a 158 GeV nucleon$^{-1}$ Indium beam at CERN \citep{Yoon2007Calib, Ahn2006BT, Marrocchesi2004, Park2004}.
The energy response was linear up to the maximum beam energy of $\sim$9 TeV.
Above the available accelerator beam energy, MC simulations indicate that
the calorimeter response is quite linear in the CREAM measurement energy range.
Simulations also indicate that the calorimeter energy resolution is nearly energy independent \citep{Ahn2001}.
Nevertheless, our energy deconvolution included corrections
for the small energy dependence of the energy resolution due to shower leakage
\citep{Ahn2009CREAM2}.

\subsection{Spectral Deconvolution}

Entries in the deposited energy bins were deconvolved into incident energy bins using matrix relations.
The counts, $N_{inc,i}$, in incident energy bin $i$ were estimated from the measured counts, $N_{dep,j}$,
in deposited energy bin $j$ by the relation \citep{Buckley1994,Ahn2006asr}
\begin{equation}
    N_{inc,\, i} = \sum_{j} P_{ij} N_{dep,\, j},
\end{equation}
where matrix element $P_{ij}$ is a probability that the events in the deposited energy bin $j$ 
are from incident energy bin $i$. 
The matrix element $P_{ij}$ 
was estimated from the response matrix generated by MC simulation results 
obtained separately for protons and helium nuclei.
The response matrix and corresponding deconvolution matrix 
were generated and tested by varying the indices between $-$2.5 and $-$2.8.
We verified that the flux deconvolution process was not sensitive to the assumed spectral index used, within that range, to generate the matrix elements.

The MC simulations for helium and heavy nuclei used FRITIOF/RQMD/DPMJET-II \citep{Kim1999,Wang2001} interfaced
to the GEANT/FLUKA 3.21 hadronic simulation package.
FRITIOF \citep{Andersson1993} is based upon semiclassical considerations of string dynamics 
for high-energy hadronic collisions.
The relativistic quantum molecular dynamics (RQMD) model was adopted for simulations of heavy ions
for energies in the center-of-mass frame less than 5 GeV nucleon$^{-1}$.
RQMD is a semiclassical microscopic approach which combines classical propagation 
with stochastic interactions \citep{Sorge1995}.
DPMJET-II \citep{Ranft1995, Ferrari1996} was based on the dual parton model, 
a framework for hadron\textendash hadron interactions and production 
in hadron\textendash nucleus and nucleus\textendash nucleus collisions at high energies.   

\subsection{Background Corrections}
\label{sec:background}
The primary background is comprised of events with misidentified charge,
which result mainly from secondary particles generated by interactions above the SCD
or from particles back-scattered from the calorimeter.
This is the case for the protons; however, there is an additional cause of misidentified events for helium nuclei:
the proton $dE/dx$ Landau tail.
Misidentified event counts of protons and helium nuclei were estimated
from the MC simulations with a power-law input spectrum. 
Due to the Landau tails, back-scattered and secondary particles,
5.1$\%$ of measured protons were misidentified helium nuclei
and 6.8$\%$ of measured helium nuclei were misidentified protons, as shown in Table \ref{tbl:effi}. 
About 0.2$\%$ of incident carbon nuclei were identified as protons,
and 2.8$\%$ of incident carbon nuclei were misidentified as helium nuclei, 
using the energy spectra of individual cosmic-rays compiled by \citet{Wiebel-Sooth1998}.
Less than 1$\%$ of trigger and reconstructed protons and helium events are from secondary particles.
Additional background comes from the events that are not within the geometry,
but which satisfy the trigger and reconstruction conditions;
they are either entering the instrument acceptance from outside the SCD area or
exiting the side of the calorimeter instead of the bottom.
According to MC simulations, this is about 3.6$\%$ and 4.0$\%$ of the selected events
for protons and helium nuclei, respectively. 
The total background was 9$\%$ for protons and 11$\%$ for helium nuclei.

\subsection{Absolute Flux}
\label{sec:absoluteflux}
The measured spectra are corrected for the instrument acceptance
as shown below to obtain the absolute flux F:
\begin{equation}
    F = \frac{dN}{dE} \frac {1}{GF ~ \varepsilon ~ T ~ \eta},
\label{eq:flux}
\end{equation}
where $dN$ is the number of events in an energy bin, $dE$ is the energy bin size, 
$GF$ is the geometry factor, $\varepsilon$ is the efficiency (defined below),
$T$ is the live time, and $\eta$ is the survival fraction 
after accounting for atmospheric attenuation. 
The geometry factor was calculated to be 0.43 m$^{2}$ sr using an MC simulations
by requiring the extrapolated calorimeter trajectory of the incident particle
to traverse the SCD active area and the bottom of the calorimeter. 
Out of 42 days of the flight, the stable period was about 24 days when no commands were sent,
e.g., for instrument tuning, power-cycle, or high-voltage adjustments.
After the dead-time correction, 
the live time, $T$, of 1,099,760 s was used for this analysis.

\noindent \textit{Efficiency.}
The efficiency, $\varepsilon$ in Equation (\ref{eq:flux}) includes efficiencies
from all analysis steps, including trigger condition, event reconstruction,
charge identification, and removing events with late interactions:
\begin{equation}
    \varepsilon = \varepsilon_{trig} ~ \varepsilon_{rec} ~ 
   \varepsilon_{sel}  ~ \varepsilon_{charge}\,. 
\end{equation}
The trigger efficiency, $\varepsilon_{trig}$, was obtained from the fraction of events 
satisfying the trigger condition among all events within the geometry,
i.e., passing through the bottom of the calorimeter and the SCD active area, using MC simulations.
This is energy dependent at low energies where the trigger is not fully efficient.
Above 3 TeV, it is nearly constant around 76$\%$ for protons and 91$\%$ for helium nuclei, respectively.
The reconstruction efficiency, $\varepsilon_{rec}$, was taken
to be the ratio of events satisfying the reconstruction and trigger conditions
to events satisfying only the trigger condition.
The reconstruction efficiency was 98$\%$ for protons and 99$\%$ for helium nuclei, respectively,
based on MC simulations.
The event selection efficiency, $\varepsilon_{sel}$,
was estimated with the MC simulations
after removing events with late interactions
and was 90$\%$ protons and 96$\%$ for helium nuclei.
The charge efficiency, $\varepsilon_{charge}$, takes into account lost events
due to the noisy or dead SCD channels, interactions above SCD and misidentified charges.
It was calculated to be 77$\%$ for protons and 67$\%$ for helium nuclei, respectively,
using MC simulations.
The efficiencies are summarized in Table \ref{tbl:effi}.

\begin{deluxetable}{lcc}
\tabletypesize{\scriptsize}
\tablecaption{Efficiencies and backgrounds for absolute flux\label{tbl:effi}}
\tablewidth{0pt}
\tablehead{ \colhead{Efficiency and Background } & \colhead{Proton (\%)} & \colhead{Helium (\%)} } 
\startdata
  Trigger efficiency            & 76 $\pm$ 2 & 91 $\pm$ 1 \\
  Reconstruction efficiency     & 98 $\pm$ 1 & 99 $\pm$ 1 \\
  Late interaction events efficiency & 90 $\pm$ 1 & 96 $\pm$ 1 \\
  Charge selection efficiency             & 77 $\pm$ 2 & 67 $\pm$ 2 \\
  Background from reconstruction        & 3.6 $\pm$ 0.1 & 4.0 $\pm$ 0.2 \\
  Background from misidentified charge  & 5.1 $\pm$ 0.2 & 6.8 $\pm$ 0.2 \\
\enddata  
\end{deluxetable}

The trigger efficiency for proton and helium nuclei cannot be estimated with the flight data,
since we do not know how many un-triggered events occurred.
However, the event selection efficiency, $\varepsilon_{sel}$, and charge efficiency, $\varepsilon_{charge}$,
were estimated in a limited way using flight data
for combined protons and helium nuclei events.
It is not as accurate as individual MC simulations 
because the composition (abundance) of the incident particles is unknown.
When the abundance ratio of protons and helium nuclei was assumed to be 1:1 and
the abundance of heavy nuclei above helium nuclei was ignored,
the combined efficiencies were 68$\%$ from the flight data and 67$\%$ for the MC simulations.

\noindent \textit{Interactions in air.}
The attenuation loss due to the atmospheric overburden, 3.9 $\pm$ 0.4 g cm$^{-2}$, 
was corrected for survival fractions of protons and helium nuclei. 
This air depth was measured by pressure sensors during the flight. 
Interaction cross sections have been measured in many fixed target experiments, 
and cross sections are known up to a few tens of GeV 
\citep{Hagen1977,Webber1990,Papini1996}.
We used the cross section formula from \citet{Hagen1977} to calculate interaction lengths
and survival fractions for protons and helium nuclei. 
The mean incident angle of 35$^{\circ}$, estimated from the flight data,
was used to estimate the losses.
The survival fraction, $\eta$, used to characterize atmospheric attenuation was 
determined to be 95$\%$ for protons and 91$\%$ for helium nuclei, respectively. 

The ratio of secondary to primary protons and helium nuclei in the atmosphere above GeV energies
has been reported \citep{Kawamura1989,Abe2003}.
\citet{Papini1996} calculated that the secondary to primary proton ratio
at an air depth of 3 g cm$^{-2}$ was less than 1$\%$ above 40 GeV,
and the secondary to primary helium nuclei ratio was 
less than 2$\%$ at 10 GeV nucleon$^{-1}$. 
Our MC simulations showed that the fraction of secondary protons and helium nuclei produced 
from carbon and iron nuclei interactions in the air was less than 1$\%$ at 10 TeV. 

\noindent \textit{Energy-bin representation.}
For the number of events ($dN$) in each energy bin with upper- and lower-energy limits,
$E_{j+1}$ and $E_{j}$, respectively ($dE = E_{j+1} - E_{j}$),
the differential flux is $dN/dE$ at $E_{m}$, where $E_{m}$ can be taken as
the arithmetic mean of $E_{j}$ and $E_{j+1}$ in logarithmic range
or else using a suitably weighted average of $E_{j}$ and $E_{j+1}$.
We also investigated an alternative procedure to determine $E_{m}$,
as suggested by \citet{Lafferty1995}:
\begin{equation}
 f(E_{m}) = \frac{1}{E_{j+1} - E_{j}} \int_{E_{j}}^{E_{j+1}} f(E) dE.
\end{equation}
For a power-law spectrum, $f(E) = A E^{-\gamma}$, $E_{m}$ can be calculated as, 
\begin{equation}
 E_{m} = \Big( \frac{E_{j+1}^{1-\gamma}-E_{j}^{1-\gamma}}{(E_{j+1}-E_{j})(1-\gamma)} \Big)^{-1/\gamma}.
\end{equation}
In this analysis, $E_{m}$ was used and the difference between $E_{m}$ and the center of the bin in logarithmic range
is less than 1$\%$.

\subsection{Uncertainties}
\label{sec:uncertainties}
The statistical uncertainty in each energy bin was estimated by the relation
$\delta N_{inc,i} = \delta (\sum_{j} P_{ij} N_{dep,j})$,
considering 68.3$\%$ the Poisson confidence interval determined by \citet{Feldman1998}.
The uncertainties were estimated by propagating uncertainties of measured entries in each bin and uncertainties of deconvolution components, $P_{ij}$, 
from MC simulations, while in the paper reported by \citet{Ahn2010CREAM12}, 
uncertainties were estimated by propagating uncertainties from measured entries with $P_{ij}$. 
This estimation gives more conservative results than the reported results.

Several sources of systematic uncertainties were identified. 
The systematic uncertainties for efficiencies and backgrounds
were estimated within each energy range to account for the energy-dependent effects determined
using MC simulations.
They are summarized in Table \ref{tbl:effi}. 
Efficiency uncertainties were about 1\textendash2$\%$ and background uncertainties were about 5$\%$.

The geometry factor uncertainty was 2$\%$ for both protons and helium nuclei; 
it was estimated with MC simulations.
The precision of estimated live-time fraction was about 3.3$\%$ and 
the accuracy of estimated dead time due to timeouts in TCD readout, which delayed processing, was about 2.6$\%$.
The overall uncertainties for the estimated live time were 4$\%$ for both protons and helium nuclei.
The systematic uncertainties for the survival fractions in the atmosphere were calculated analytically.
The $p$\textendash$p$ cross section difference between 10 TeV and 100 TeV is about 28$\%$, 
according to the most recent reference from the Particle Data Group \citep{PDG2008}. 
Using a conservative estimate of 30$\%$ for cross section uncertainties,
the estimated uncertainties of survival fractions were 2$\%$ and 3$\%$
for protons and helium nuclei, respectively.
The range of incident angle was from 0$^{\circ}$ to 66$^{\circ}$.
The uncertainty in correcting for atmospheric losses introduced by using an assumed mean incident angle
was at the level of 1$\%$ for protons and 1.6$\%$ for helium nuclei.
The energy calibration accuracy was found to be 1$\%$.
The systematic uncertainties of the measured number in each energy bin,
considering the 1$\%$ energy calibration accuracy, were 3$\%$ for both protons and helium nuclei.
To estimate uncertainties in the spectral deconvolution, the unfolding procedure was repeated
by varying input spectral indices.
The difference of the proton fluxes varying the input spectral indices between 2.64 and 2.68 was 
less than 1$\%$.
Similarly, the helium flux difference was also less than 1$\%$ for input spectra between indices 2.56 and 2.60.
The overall systematic uncertainties were found to be 9$\%$ for both protons and helium nuclei. 
These systematic uncertainties are energy independent.
They do not change the spectral shape, but they might shift the normalization of the spectra up or down. 

\section{Results}
The measured proton fluxes from 2.5 TeV to 250 TeV and 
helium fluxes from 630 GeV nucleon$^{-1}$ to 63 TeV nucleon$^{-1}$ 
at the top of the atmosphere are given in Tables \ref{tbl:proton} and \ref{tbl:helium},
while previously reported results in the paper by \citet{Ahn2010CREAM12} are presented in a plot. 
The statistical uncertainties were re-estimated,
as discussed in Section \ref{sec:uncertainties}.
The CREAM proton and helium spectra are each consistent with a single power law 
over the measured range.
The best-fit parameters for the spectra for protons and helium nuclei are represented by
\begin{equation}
    \frac{d\Phi}{dE} \,=\, \Phi_{0} E^{-\beta} ~~~  (\textmd{m}^{2}\,\textmd{sr\,s\,GeV\,nucleon}^{-1})^{-1}.
\end{equation}
The best-fit parameters for the spectra for protons and helium nuclei are given by \citep{Ahn2010CREAM12}:
\begin{equation}
    \Phi_{0,p} \,=\, ( 7.8 \pm 1.9 ) \times 10^{3} 
~~(\textmd{m}^{2}\,\textmd{sr\,s})^{-1} (\textmd{GeV nucleon}^{-1})^{1.66},
\end{equation}
\begin{equation}
    \beta_{p} \,=\, 2.66 \pm 0.02, 
\end{equation}
and
\begin{equation}
    \Phi_{0,He} \,=\, ( 4.2 \pm 0.8 ) \times 10^{2} 
~~(\textmd{m}^{2}\,\textmd{sr\,s})^{-1}\,(\textmd{GeV\,nucleon}^{-1})^{1.58},
\end{equation}
\begin{equation}
    \beta_{He} \,=\, 2.58 \pm 0.02. 
\end{equation}

The spectral indices for proton and helium nuclei were calculated both with the least squares fit
and maximum likelihood method. The results from both methods were consistent.
Uncertainties for the spectral indices were estimated with the maximum likelihood method.

\begin{deluxetable}{cc}
\tabletypesize{\scriptsize}
\tablecaption{Proton differential flux measured with CREAM\label{tbl:proton}}
\tablewidth{0pt}
\tablehead{ \colhead{ Energy Bin Range} & \colhead{Flux  $\pm$ Stat. } \\ 
\colhead{ (\textmd{GeV})} & \colhead{ \textmd{(m$^{2}$~sr~s~GeV)}$^{-1}$  } }

\startdata
$2.5 \tten ^{3} - 4.0 \tten ^{3}$ 	 & $ ( 3.72 \pm 0.10 ) \tten ^{-6}$ \\
$4.0 \tten ^{3} - 6.3 \tten ^{3}$ 	 & $ ( 1.10 \pm 0.04 ) \tten ^{-6}$ \\
$6.3 \tten ^{3} - 1.0 \tten ^{4}$ 	 & $ ( 3.19 \pm 0.19 ) \tten ^{-7}$ \\
$1.0 \tten ^{4} - 1.6 \tten ^{4}$ 	 & $ ( 9.47 \pm 0.80 ) \tten ^{-8}$ \\
$1.6 \tten ^{4} - 2.5 \tten ^{4}$ 	 & $ ( 2.80 \pm 0.35 ) \tten ^{-8}$ \\
$2.5 \tten ^{4} - 4.0 \tten ^{4}$ 	 & $ ( 8.1  \pm 1.5 ) \tten ^{-9}$ \\
$4.0 \tten ^{4} - 6.3 \tten ^{4}$ 	 & $ ( 2.2  \pm 0.6 ) \tten ^{-9}$ \\
$6.3 \tten ^{4} - 1.0 \tten ^{5}$ 	 & $ ( 6.1 ^{+2.6}_{-2.2} ) \tten ^{-10}$ \\
$1.0 \tten ^{5} - 1.6 \tten ^{5}$ 	 & $ ( 1.8 ^{+1.2}_{-0.9} ) \tten ^{-10}$ \\
$1.6 \tten ^{5} - 2.5 \tten ^{5}$ 	 & $ ( 4.2 ^{+5.4}_{-3.4} ) \tten ^{-11}$ \\
\enddata  
\end{deluxetable}

\begin{deluxetable}{cc}
\tabletypesize{\scriptsize}

\tablecaption{Helium differential flux measured with CREAM\label{tbl:helium}}
\tablewidth{0pt}
\tablehead{ \colhead{Energy Bin Range } & \colhead{Flux $\pm$ Stat. } \\
\colhead{\textmd {(GeV nucleon$^{-1}$)}} & \colhead{ (\textmd{m$^{2}$~sr~s~GeV~nucleon$^{-1}$})$^{-1}$ }}

\startdata
$6.3 \tten ^{2} - 1.0 \tten ^{3}$ 	 & $ ( 1.42 \pm 0.04 ) \tten ^{-5}$ \\
$1.0 \tten ^{3} - 1.6 \tten ^{3}$ 	 & $ ( 4.35 \pm 0.16 ) \tten ^{-6}$ \\
$1.6 \tten ^{3} - 2.5 \tten ^{3}$ 	 & $ ( 1.31 \pm 0.07 ) \tten ^{-6}$ \\
$2.5 \tten ^{3} - 4.0 \tten ^{3}$ 	 & $ ( 3.83 \pm 0.31 ) \tten ^{-7}$ \\
$4.0 \tten ^{3} - 6.3 \tten ^{3}$ 	 & $ ( 1.27 \pm 0.14 ) \tten ^{-7}$ \\
$6.3 \tten ^{3} - 1.0 \tten ^{4}$ 	 & $ ( 4.19 \pm 0.64 ) \tten ^{-8}$ \\
$1.0 \tten ^{4} - 1.6 \tten ^{4}$ 	 & $ ( 1.15 \pm 0.27 ) \tten ^{-8}$ \\
$1.6 \tten ^{4} - 2.5 \tten ^{4}$ 	 & $ ( 3.4 ^{+1.1}_{-1.0} ) \tten ^{-9}$ \\
$2.5 \tten ^{4} - 4.0 \tten ^{4}$ 	 & $ ( 8.2 ^{+4.9}_{-3.8} ) \tten ^{-10}$ \\
$4.0 \tten ^{4} - 6.3 \tten ^{4}$ 	 & $ ( 2.9 ^{+2.4}_{-1.5} ) \tten ^{-10}$ \\
\enddata
\end{deluxetable}

The CREAM proton spectrum is harder than previous measurements at lower energies such as
AMS \citep{AMS2002}, 2.78 $\pm$ 0.009 at 10\textendash200 GV and BESS \citep{Haino2004},
2.732 $\pm$ 0.011 from 30 GeV to a few hundred GeV. 
Likewise, the CREAM helium spectrum is harder than AMS, 2.740 $\pm$ 0.01 at 20$-$200 GV and BESS,
2.699 $\pm$ 0.040 from 20 GeV nucleon$^{-1}$ to a few hundred GeV nucleon$^{-1}$. 

\begin{figure} [t]
\epsscale{.65} 
\plotone{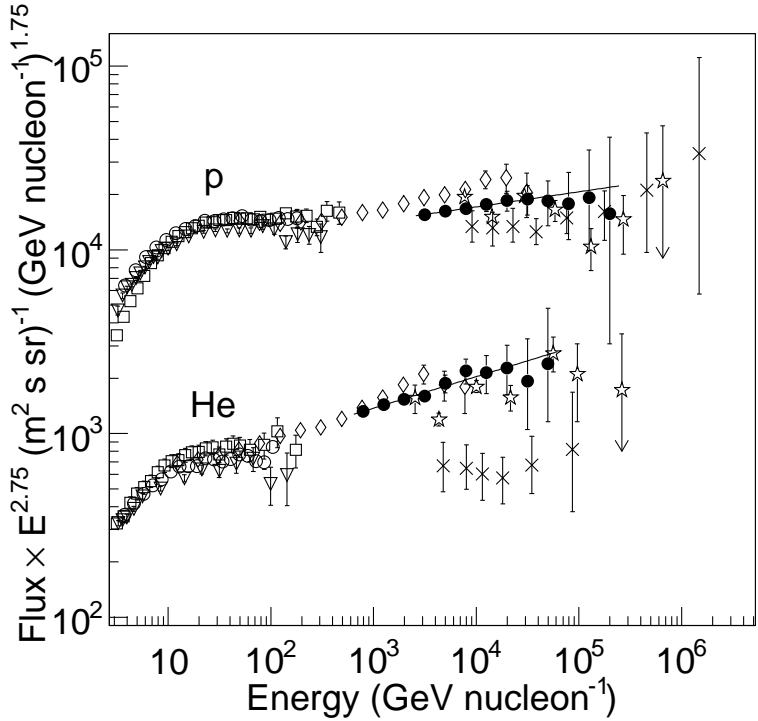}
\caption{CREAM Proton and helium differential $Flux \cdot E^{2.75}$ in GeV nucleon$^{-1}$ 
at the top of the atmosphere.
The CREAM proton and helium spectra (filled circles) are 
shown together with previous measurements:
BESS (squares), 
CAPRICE98 (downward triangles), 
AMS (open circles),
ATIC-2 (diamonds), 
JACEE (stars)
and RUNJOB (crosses).
The lines represent power-law fits with spectral indices of -2.66 $\pm$ 0.02 for protons and 
-2.58 $\pm$ 0.02 for helium nuclei, respectively. 
\label{fig:cream1hhe275}}
\end{figure}

Figure \ref{fig:cream1hhe275} compares our measured spectra with previous measurements:
AMS, BESS, CAPRICE98 \citep{Boezio2003}, ATIC-2 \citep{Panov2009}, JACEE \citep{Asakimori1998}
and RUNJOB \citep{Derbina2005}. 
The error bars shown in the figures represent the statistical uncertainties.
The CREAM results are consistent with JACEE where its measurement energy range overlaps
with CREAM but indicate higher fluxes, particularly for helium, with respect to RUNJOB.
The proton and helium fluxes are both higher than that expected 
by extrapolating the power law fitted to the lower-energy measurements,
which verifies that our TeV spectra are harder
than the lower-energy spectra. 
At 20 TeV nucleon$^{-1}$ the helium flux measured by CREAM is
about 4$\sigma$ higher than the flux expected
from a power-law extrapolation of the AMS helium flux and spectral index.

\begin{figure} [t]
\epsscale{.65} 
\plotone{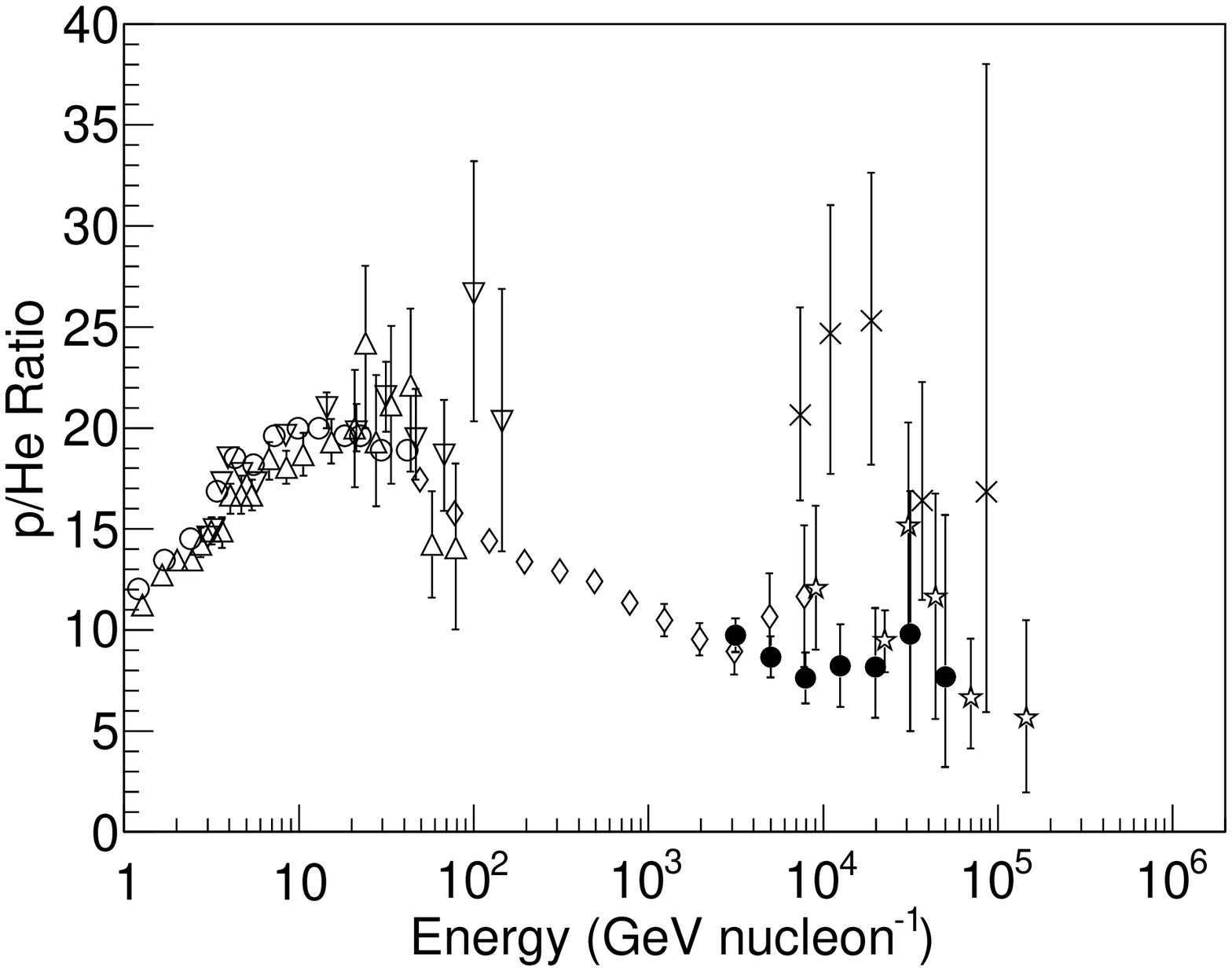}
\caption{Ratio of protons to helium nuclei as a function of energy in GeV nucleon$^{-1}$. 
The CREAM (filled circles) ratio of proton to helium is compared with previous measurements:
ATIC-2 (diamonds),
CAPRICE94 (upward triangles), 
CAPRICE98 (downward triangles), 
LEAP (open circles),
JACEE (stars),
and RUNJOB (crosses).
\label{fig:cream1ratio}}
\end{figure}

The proton to helium ratio as a function of energy provides insight into 
whether the proton and helium spectra have the same spectral index.
This has long been a tantalizing question, 
mainly because of the limited energy range individual experiments could cover. 
The ratio from the first CREAM flight provides a much needed higher energy, low-statistical uncertainty, measurement.
The ratio is compared with previous measurements in Figure \ref{fig:cream1ratio}:
ATIC-2, CAPRICE94 \citep{Boezio1999}, CAPRICE98, JACEE \citep{Asakimori1993he}, 
LEAP \citep{Seo1991}, and RUNJOB.
The CREAM ratios are consistent with JACEE where its measurement energy range overlaps.
The measured CREAM ratio at the top of the atmosphere is on average 9.1 $\pm$ 0.5
for the range from 2.5 TeV nucleon$^{-1}$ to 63 TeV nucleon$^{-1}$,
which is significantly lower than the ratio of $\sim$20 obtained
from the lower-energy measurements.

\section{Discussion and Conclusion}

The energy spectra of primary cosmic rays are known with good precision
up to energies around 10$^{11}$ eV, where magnetic spectrometers
have been able to carry out such measurements.
Above this energy the composition and energy spectra are not accurately known,
although there have been some pioneering measurements \citep{Mueller1991, Asakimori1998, Apanasenko2001}.
The collecting power of CREAM is about a factor of two larger than
that of ATIC for protons and helium nuclei and,
considering the much larger geometry factor of the TRD, about a factor of 10 larger for heavier nuclei. 
TRACER has a larger geometry factor than CREAM, but a smaller dynamic charge range ($Z$ = 8\textendash26)
was reported for its 10-day Antarctic flight. 
Although its dynamic charge range was improved to $Z$ =3\textendash26 for its $\sim$4 day flight
from Sweden to Canada in 2006, it is still insensitive to protons and helium nuclei.

The CREAM payload maintained a high altitude, 
corresponding to an atmospheric overburden of 3.9 g cm$^{-2}$ for vertically incident particles.
That implies about 6.8 g cm$^{-2}$ at the maximum acceptance angle for this analysis,
which is smallest among comparable experiments. 
For example, the average vertical depth for RUNJOB was more than twice that of CREAM,
due to its low flight altitude. 
Considering the RUNJOB acceptance of particles at large zenith angles,
its effective atmospheric depth was as large as 50 g cm$^{-2}$. 
For that depth, large corrections are required to account for the fact that 41$\%$ of protons
would have interacted before reaching the detector.

The CREAM calorimeter is much deeper than either that of JACEE or RUNJOB,
so it provides better energy measurements.
CREAM also has excellent charge resolution,
sufficient to clearly identify individual nuclei,
whereas JACEE and RUNJOB reported elemental groups.
Our observation did not confirm a softer spectrum of protons above 2 TeV reported by \citet{Grigorov1970}
or a bend around 40 TeV \citep{Asakimori1993p}.
An increase in the flux of helium relative to protons could be interpreted
as evidence for two different types of sources for protons and helium nuclei
as proposed by \citet{Biermann1993}. 
The observed harder spectra compared to prior low-energy measurements
may require a significant modification of conventional acceleration and propagation models,
with significant impact for the interpretation of other experimental observations.  

The CREAM experiment was planned for Ultra Long Duration Balloon (ULDB) flights
lasting about 100 days with super-pressure balloons.
While waiting for development of these exceptionally long flights,
the CREAM instrument has flown five times on LDB flights in Antarctica. 
It should be noted that a 7 million cubic foot ($\sim$0.2 million cubic meters)
super-pressure balloon was flown successfully for 54 days during the 2008-2009 austral summer season.
As ULDB flights become available for large science payloads,
long-duration exposures can be achieved faster and more efficiently
with reduced payload refurbishment and launch efforts.
Whatever the flight duration, data from each flight reduces the statistical uncertainties
and extends the reach of measurements to energies higher than previously possible. 

\acknowledgments

This work was supported in the U.S. by NASA grants NNX07AN54H, NNX08AC11G,
NNX08AC15G, NNX08AC16G and their predecessor grants,
in Italy by INFN, in Korea by the Creative Research Initiatives of MEST/NRF.
The authors wish to acknowledge NASA/WFF for provision and operation of flight support systems;
CERN for provision of excellent accelerator beams;
and CSBF,
National Science Foundation's Office of Polar Programs,
and Raytheon Polar Services Company for outstanding support of launch,
flight, and recovery operations in Antarctica.

\clearpage

\end{document}